# Advection kinetics induced self–assembly of colloidal nanoflakes into microscale floral structures

**Purbarun Dhar** *,#

Department of Mechanical Engineering, Indian Institute of Technology Ropar,

Rupnagar–140001, India

Department of Mechanical Engineering, Indian Institute of Technology Madras,

Chennai–600036, India

**\*E–mail**: purbarun@iitrpr.ac.in ; pdhar1990@gmail.com

**Tel**: +91-1881-24-2119

# *Parts of the research work was done during the author's previous affiliation to IIT Madras and partly during affiliation to IIT Ropar*

## Abstract

This article explores the governing role of the internal hydrodynamics and advective transport within sessile colloidal droplets on the self-assembly of nanostructures to form floral patterns. Water-acetone binary fluid and $Bi_2O_3$ nanoflakes based complex fluids are experimented with. Micro-liter sessile droplets are allowed to vaporize and the dry-out patterns are examined using scanning electron microscopy. The presence of distributed self-assembled rose like structures is observed. The population density, structure and shape of the floral structures are noted to be dependent on the binary fluid composition and nanomaterial concentration. Detailed microscopic particle image velocimetry analysis is undertaken to qualitatively and quantitatively describe the solutal Marangoni advection within the evaporating droplets. It has been shown that the kinetics, regime and location of the internal advection are responsible factors towards the hydrodynamics influenced clustering, aggregation and self-assembly of the nanoflakes. In addition, the size of the nanostructures and the viscous character of the complex fluid also play dominant roles. The resulting interplay of hydrodynamic behavior, adhesion and cohesion forces during the droplet dry-out phase, and thermodynamic energy minimization leads to formation of such floral structures. The findings may have strong implications towards modulating micro-hydrodynamics induced self-assembly in complex fluids.

## 1. Introduction

Evaporative self-assembly and pattern formation in colloidal complex fluid droplets has been an area of academic interest and applications. These include design and development of super-lattices [1, 2], engineering building proteins of life [3], nano-biotechnology [4], nanoscale origami and structure modulation [5, 6], crystal engineering for optics and photonics [7], supra-molecular drugs [8, 9], nanoscale lithography [10], nematic crystals [11], and so on. The pattern and structure assembly is typically governed by micro-hydrodynamics forces in conjunction with physico-chemical interactions, such as adhesion-cohesion [12], electro-magneto-thermal interactions [13, 14] and/or chemical reactions at the microscale [15]. Consequently, the role of micro-hydrodynamics and advection kinetics within such complex fluid droplets and the subsequent dry-out behavior that leads to such assembly required thorough understanding. Recently, some aspects of evaporative transport [16], internal hydrodynamics [17] and Marangoni advection [18] is complex colloidal droplets have been explored. The modulation of such micro-hydrodynamics employing opto-thermal [19], electric [20, 21] and magnetic [22, 23] fields have also been discussed in literature.

The present article discusses the role of the internal Marangoni advection and its kinetics in modulating the physical self-assembly of nanostructures. Binary fluids [24] of water and acetone are employed to modulate the characteristics of the advection hydrodynamics during evaporation. Bismuth oxide ($Bi_2O_3$) nanoflakes are used as the colloidal phase and their self-assembly is studied using high resolution scanning electron microscopy (HRSEM). Physical clustering, agglomeration and packing assembly into microscale floral (rose) shapes are noted under different conditions. The number density and conformational characteristics of the flowers are explained based on the internal hydrodynamics. Microscopic particle image velocimetry (PIV) is used to qualitatively and quantitatively understand the Marangoni advection kinetics. The strength of micro-hydrodynamics, droplet lifetime and colloidal concentration are explored and self-assembly behavior is explained. The study may have strong implications on developing microscale hydrodynamics modulated colloidal patterning and assembly strategies.

## 2. Experimental materials and methodologies

The experimental details are described briefly (details available elsewhere [18, 25]). For the HRSEM (high resolution scanning electron microscopy) imaging, aluminum foil cut-outs (10 mm x 10 mm) are used as conducting substrates. The foils are cautiously (to prevent wrinkling and crumbling) cleaned with acetone and oven dried overnight. Similarly processed sterile glass slides are used as the substrates for the PIV. The wetting behavior of both processed substrates are very similar (contact angle for water droplet range from 32–37º, and evaporating droplet lifetime for water is within 5-7 % and acetone droplets are within 2-3 %). Deionized water and acetone (>99.5% pure, AR, Sigma Aldrich) are used as test fluids. The $Bi_2O_3$ nanoflakes (fig. S1, supporting information SI [a], Alfa Aeser, India) are dispersed in the base fluids by ultrasonic disruption (Oscar Ultrasonics, India). Colloid concentrations of 0.005, 0.01, 0.025, 0.05 and 0.1 wt. % are studied. The base fluids are different proportions of water (W) and acetone (A), viz. W100 A0, W80 A20, W60 A40, W40 A60, W20 A80, and W0 A100 (all wt./wt. %).

A precision microliter chromatography syringe (Holmarc Opto-mechatronics, India) is used to dispense a 10 µL droplet on the foils, and allowed to evaporate within an enclosed acrylic chamber (temperature and relative humidity conditions 30±2 ºC and 48±4 % for all experiments). The evaporation is recorded at 10 fps using a CCD camera (Holmarc Opto-mechatronics, India) with long-distance-microscope lens, and droplet life-times are assessed. The nanomaterial deposits are imaged using HRSEM. For quantification of the internal advection, the droplets are placed on glass slides and observed under a CMOS camera fit microscope (Holmarc Opto-mechatronics, India). The fluids are seeded with hydrophilic, 500 nm carbon particles for PIV. The microscope is focused at the horizontal mid-plane of the droplet and imaged at 30 fps with bright-field illumination. The images are post-processed in ImageJ [26], and converted to color inverted binary, to transform the seed particles to white with the droplet to black. The velocimetry analysis is done in the open source code PIV-lab [27]. 1000 consecutive images are processed using 4 pass cross-correlation algorithm. Noise suppression and contrast enhancement algorithms are used to obtain better peak matching and high SNR. Detailed description of the PIV methodology followed is reported by the author [25].

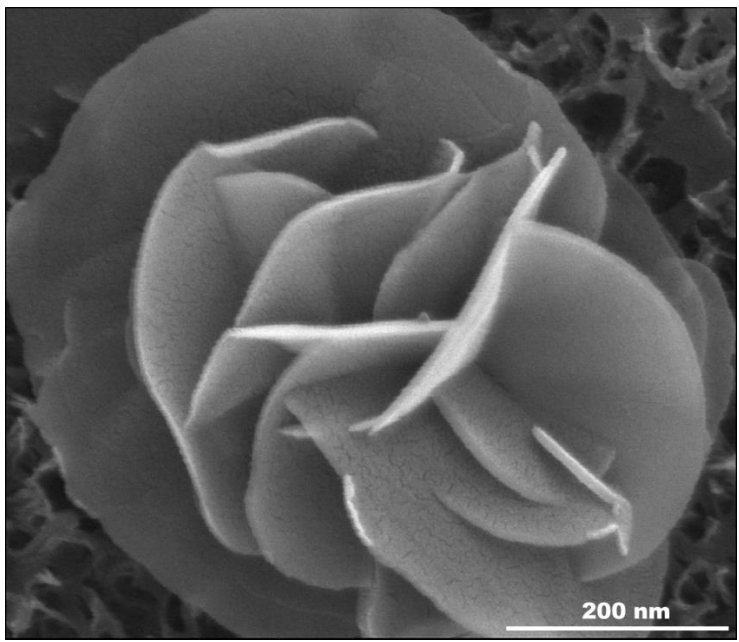

**Figure 1:** HRSEM image of a typical microscale rose structure by individual nanoflakes formed post evaporation of a water-acetone-$Bi_2O_3$ (40% water, 60% acetone, and 0.025 wt. % nanomaterials) colloid droplet.

## 3. Results and discussions

The colloidal patterns on Al foils are examined using HRSEM, and dried bed of nano-flakes is noted. Under certain conditions (discussed subsequently), the pattern is characterized by numerous localized distribution of microscale floral assemblies, typically composed of the larger nanoflakes. An illustration of a typical self-assembled 'rose' pattern is in fig. 1. The smaller scale flakes form the bed of the deposition, whereas the larger flakes adhere by cohesion to form the floral structures. The bed deposition is formed by the 40-60 nm flakes (refer fig S1, SI) and the floral patterns are formed by the 200-400 nm flakes. The mean overall size (base diameter) of such self-assembled flowers ranges between 500-1000 nm (fig. 1 and 2a). The content of the acetone in the binary fluid colloid governs the symmetry of the assembled structures. Near the observed optimal water-acetone proportion (discussed later), the rose formations are majorly noticed to possess good symmetry (fig. 1). On the contrary, acetone content displaced from the determined optima is noted to yield rose structures with mostly asymmetric topology (fig. 2a), with additional clustering or fusion of two or more flowers.

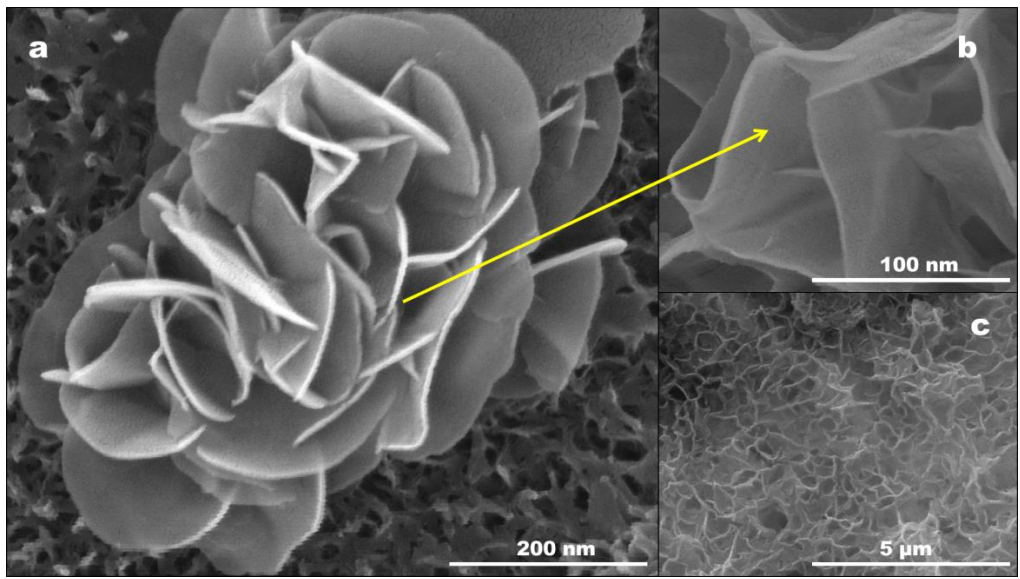

**Figure 2: (a)** HRSEM image of a microscale rose structure from a water-acetone-$Bi_2O_3$ colloid (60% water, 40% acetone, and 0.025 wt. % nanomaterials) droplet **(b)** zoom view within the structure showing the layering of the individual flakes **(c)** formation of random layers of nanoflakes for only water or acetone based colloids.

The formation of the flowers is due to the interplay of hydrodynamic and cohesive-adhesive forces between the larger nanoflakes and the base fluid during evaporative drying. This forms weakly structured flakes (fig. 2b), however, the cohesive forces between the flakes is strong enough to withstand damage to the structural integrity of the flowers during the gold film deposition before HRSEM analysis. The experiments were designed using proper parametric variation to establish facts from artefacts. The results of the parametric experimentation are illustrated in fig. 3. The dried patterns from both pure water (W100 A0) and acetone (W0 A100) based colloids (irrespective of nanomaterial concentration) do not show any sign of floral pattern formation (fig. 3a). The SEM images only show uniformly distributed bed of the nanostructures (fig. 2 c). This clearly indicates that the self-assembly phenomena is driven by the internal advection generated within the binary fluid droplet. This further brings to the forefront that unequal evaporation generated concentration and/or thermal gradient driven advection within the droplet [16-18] is a major responsible phenomena for the observed structures. Based on the population count of such structures from microscopy, typical number density plot has been shown (fig. 3a).

The highest number density of such floral structures is consistently observed for the W60 A40 fluid, and the structures possess good symmetry. The immediate neighboring mixtures also yield such structures, albeit reduced in number density, and asymmetric defects are innate in such structures. From the nature of the plot it is evident that addition of acetone to water is a governing mechanism that induces the self-assembly of the larger flakes. However, beyond certain acetone content, reversal of trend is consistently prominent, which furthers the proposition of thermo-solutal Marangoni advection as a governing mechanism. Fig. 3b illustrates the influence of the nanoflakes concentration on the number density of the floral structures. Viscosity of the colloids (refer SI, measured using capillary viscometry [28, 29], Anton Paar) illustrate that the viscosity enhances suddenly beyond the ~0.025 wt. % concentration. Consequently, proposition can be placed that in the dilute regime (<0.025 wt. %), the number density of the flakes is not sufficient to form well defined and numerous floral structures, whereas beyond the observed optima concentration, the enhanced viscosity of the colloid opposes the internal advection caliber, which plausibly leads to reduction in formation of self-assembled microstructures.

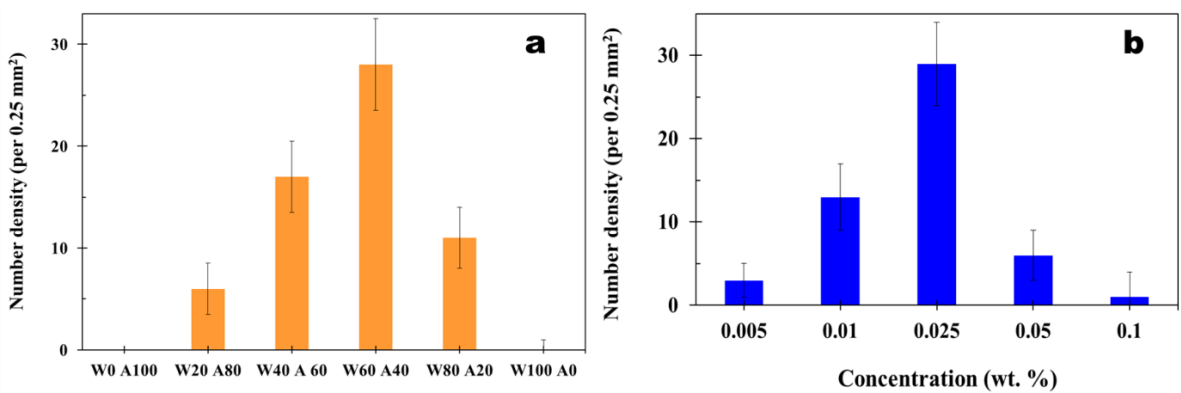

**Figure 3: (a)** Number density (per 0.25 mm$^2$) of the rose structures for different evaporating colloidal droplet binary fluid content (at 0.025 wt. % Bi$_2$O$_3$) **(b)** number density (per 0.25 mm$^2$) of the rose structures for different Bi$_2$O$_3$ content at 60-40 water acetone composition.

With evidence that internal advection results in the self-assembly, the kinetics of the advection and the observations in fig. 3 need to be correlated. The temporally averaged advection velocity during the initial phases of evaporation (within few seconds of placing the droplet) and that during the droplet half-life period (refer fig. 4b for the non-dimensional droplet life-times) are quantified using microscopic PIV. The velocities are illustrated in fig. 4a. The initial regime velocities are observed to enhance with addition of acetone to water, and is due to solutal Marangoni convection and solute gradient advection [17, 18] within the binary fluid droplet. Since acetone vaporizes faster compared to the water phase (fig. 4b), the solutal advection kinetics is a function of time. The advection velocity at the half-life (fig. 4a) shows that the velocities decay significantly, except for the case of pure acetone and W20 A80. The velocity at half-life for the optimal case of W60 A40 is interestingly similar to the water case, despite the initial regimes being largely dominated by advection.

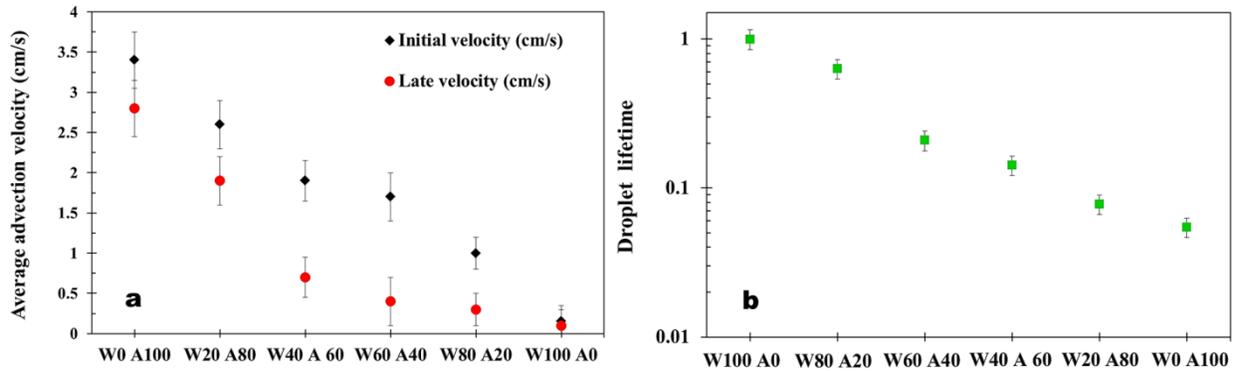

**Figure 4: (a)** Average internal advection velocity within different binary fluid droplets (with 0.025 % particles). Initial velocity corresponds to the very initial regime, while later regime corresponds to the half-life regime of the droplet **(b)** Normalized droplet lifetime for different fluids (with 0.025 % particles). The droplet lifetimes are normalized with the lifetime of the water droplet.

Hence, the initial and late regime advection characteristics both govern the floral pattern formation. A picture of the internal kinetics can now be proposed. The 40-60 nm flakes behave as perfect Stokesian bodies and follow the advection behavior, and are deposited evenly along the substrate during the dry-out process [30]. The larger flakes however, due to their greater adhesion-cohesion interplay, are deposited in agglomerates, as local structures or clusters during the dry-out. While the initial regime advection currents (fig. 5 b1) allow the clustering of the larger flakes, the fast advection and mixing motion in the W0 A100 and W20 A80 even in the late regime (fig. 5 b2) prevents the clusters from stabilizing structurally during the dry-out. On the contrary, the initial advection in the W60 A40 (fig. 5 a1) is potent enough to support the localized clustering and aggregation. The later advection decays substantially (fig. 5 a2), which permits for the aggregates to stabilize structurally under the balance of adhesion-cohesion during the dry-out phase, and form the rose like shapes, to achieve thermodynamically favorable conformations [31-33]. In the low acetone concentrations, the initial advection itself is weak and cannot induce the favorable clustering. Further details on the internal advection regimes and periods have been discussed in fig. S3 (refer SI).

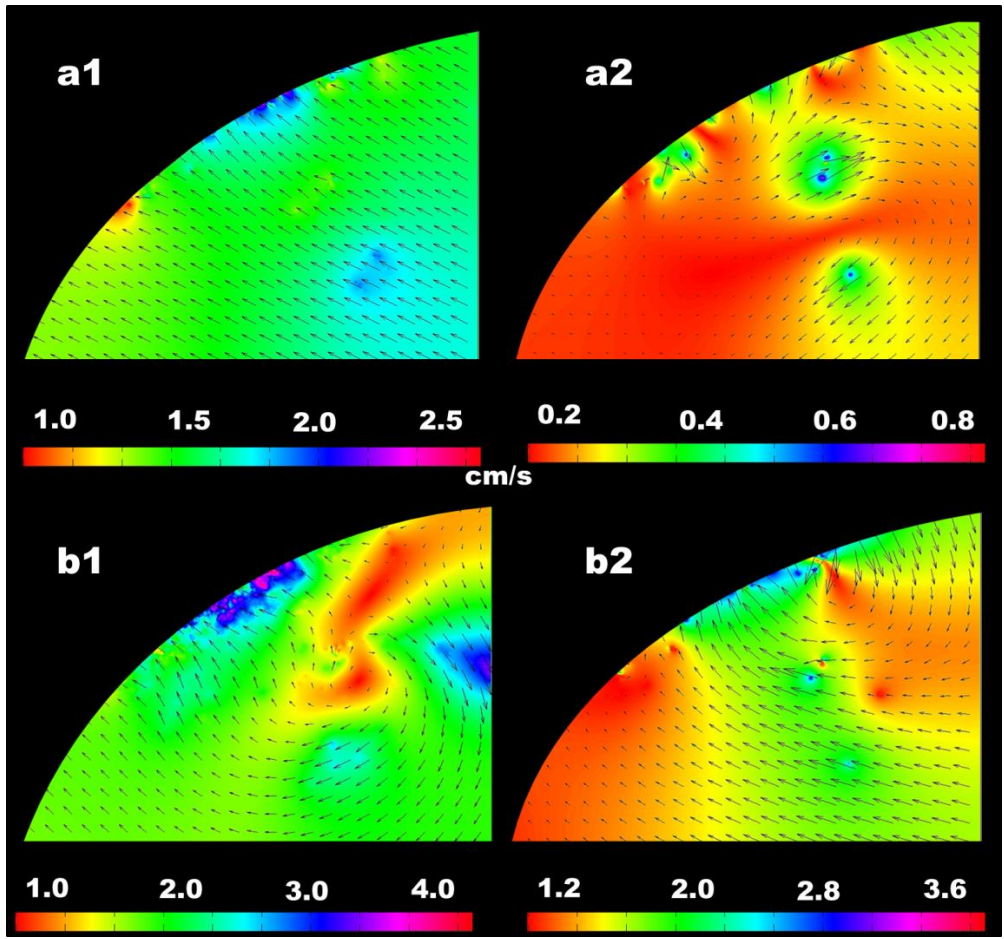

**Figure 5:** Post-processed velocity contours and vector field at the droplet mid-plane (horizontal) **(a1)** W60 A40 sample at initial regime **(a2)** W60 A40 sample at late regime **(b1)** W0 A100 sample at initial regime **(b2)** W0 A100 sample at initial regime. The contours and velocity fields correspond to the temporal average of 1000 velocimetry frames (frame rate 30).

## 4. Summary and conclusions

To infer, the article explores the role played by internal Marangoni advection and the hydrodynamic characteristics within sessile droplets of complex fluids on formation of microscale floral structures. $Bi_2O_3$ nanoflakes based water-acetone binary fluids are tested. Evaporation induced clustering and self-assembly of the flakes to form rose shaped structures is noted from HRSEM. It is seen that the water-acetone proportions acutely govern the population density of such structures. Microscopic PIV is done to understand the kinetics of the internal hydrodynamics and Marangoni advection, and it is revealed that the strength, nature and spatial extent of the advection are the governing factor. Likewise, the size of the nanoflakes and viscosity of the colloid also play dominant roles. The findings are first of its kind and directly correlates the internal micro-hydrodynamics within a droplet to the observed colloidal self-assembly and structuring.

**Acknowledgements:** The author acknowledges the Sophisticated Analytical Instrumentation Facility (SAIF) of IIT Madras for the HRSEM facilities. The author also acknowledges the funding by IIT Ropar (grant number IITRPR/Research/193).

**Conflict of Interest:** The author has no conflicts of interest with any individual or agency.